\documentclass[11pt,twoside]{article}

\newcommand\nodata{ ~$\cdots$~ }

\usepackage{asp2006}
\usepackage{epsf}
\usepackage{psfig}
\usepackage{lscape}

\def\plotfiddle#1#2#3#4#5#6#7{\centering \leavevmode
\vbox to#2{\rule{0pt}{#2}}
\includegraphics{#1}}
\begin{document}

\title{Deconstructing Disk Velocity Distribution Functions in the Disk-Mass Survey}   
\author{Kyle B. Westfall\altaffilmark{1}, Matthew
A. Bershady\altaffilmark{1}, Marc A. W. Verheijen\altaffilmark{2},
David R. Andersen\altaffilmark{3}, \& Rob A. Swaters\altaffilmark{4}}
\altaffiltext{1}{Dept. of Astronomy, U. of Wisconsin, 475 N. Charter St., Madison, WI\ \ 53706, USA}
\altaffiltext{2}{Kapteyn Inst., Landleven 12, 9747 AD Groningen,
the Netherlands}
\altaffiltext{3}{NRC Herzberg Inst. of Astrophysics, 5071 W.
Saanich Road, Victoria, BC, Canada\ \ V9E 2E7}
\altaffiltext{4}{Dept. of Astronomy, U. of Maryland,
College Park, MD\ \ 20742, USA}

\begin{abstract} 

We analyze integral-field ionized gas and stellar line-of-sight
kinematics in the context of determining the stellar velocity
ellipsoid for spiral galaxies observed by the Disk-Mass Survey. Our
new methodology enables us to measure, for the first time, a radial
gradient in the ellipsoid ratio $\sigma_z/\sigma_R$. Random errors
in this decomposition are 15\% at two disk scale-lengths.

\end{abstract}

The stellar velocity ellipsoid describes the velocity
distribution of positionally coincident stars in galaxy disks via an
imaginary surface outlining the 3D velocity dispersion about their
mean orbit.  Its axis magnitudes and axial ratios provide physical
insight to disk stability, disk mass surface density, and disk heating
mechanisms.  Development of a robust method for decomposing the
line-of-sight velocity distribution (LOSVD) into the
$R$, $\phi$, and $z$ components is, therefore, critical for reliable
physical diagnostics of galaxy disks. Furthermore, uncertainties in the
ellipsoid axial ratios are among the two largest contributors to the
Disk-Mass Survey (DMS) error-budget -- see Verheijen et al. (2007) for the DMS description. The dominant
component, empirical conversion of radial scale-lengths to vertical
scale-heights, contributes $<$25\% uncertainty in $M/L$ zero points per
galaxy, based on analysis of independent edge-on samples
(e.g., Schwarzkopf \& Dettmar 2000, Kregel et al. 2002). The
ellipsoid, however, can be constrained internally for each DMS galaxy.
As shown in Figure 1, an ellipsoid-deprojection error of
10\% contributes a measurement uncertainty in $M/L$ of $\sim 15$\% 
for typical galaxy inclinations in our
sample.  Given the astrophysical import of the ellipsoid, we focus
intensively on its measurement here.

We have previously reported ellipsoid decompositions
for NGC 3949 and NGC 3982 using a method focusing only on data
within 40$^{\circ}$ of the major axis (Westfall et al. 2007).  (These results were compared with Shapiro et al. (2003) who closed the deprojection equations via
major- and minor-axis long-slit observations and the Epicycle
Approximation [EA].)  By assuming both EA and the Asymmetric Drift (AD) equations held, we
calculated $\sigma_R$ via a simplified AD equation using only the gas and stellar rotation
curves.
The measurement assumed the gas ([{\scshape Oiii}]$\lambda 5007$\AA) rotation
follows the circular speed, the e-folding length of $\sigma_R$ was 2$h_R$
(twice the disk photometric scale
length), and the ellipsoid remains oriented with the cylindrical coordinates
at $z\neq0$ with constant anisotropy.  The full ellipsoid model was then
decomposed using only $\eta\equiv\sigma_z/\sigma_R$ as a
parameter, and tested against $\sigma_{\rm maj}$.  This highly non-parametric
approach likely produced systematically high $\sigma_z/\sigma_R$ ratios
by smearing the natural sin$^2\phi$ function of the
LOS velocity dispersion, $\sigma_{\rm LOS}$,
onto the 40$^{\circ}$ ``major axis'' wedge.  In our
current analysis: (1) We model both radial and
azimuthal variations in the full 2D LOSVD kinematics simultaneously; (2) We
free the e-folding length of the dispersion (previously fixed at 2$h_R$); and (3) We isolate EA- and AD-only decomposition methodologies. Table 1 provides the
kinematic geometry and rotation
curve parameters derived from our gas ($g$) and stellar ($\star$) velocity
fields.  We
compare results from our current analysis with our previous report in
Westfall et al. (2007) in Table 2. The fitted data are presented in
Figure 2.

\begin{figure}
\plotfiddle{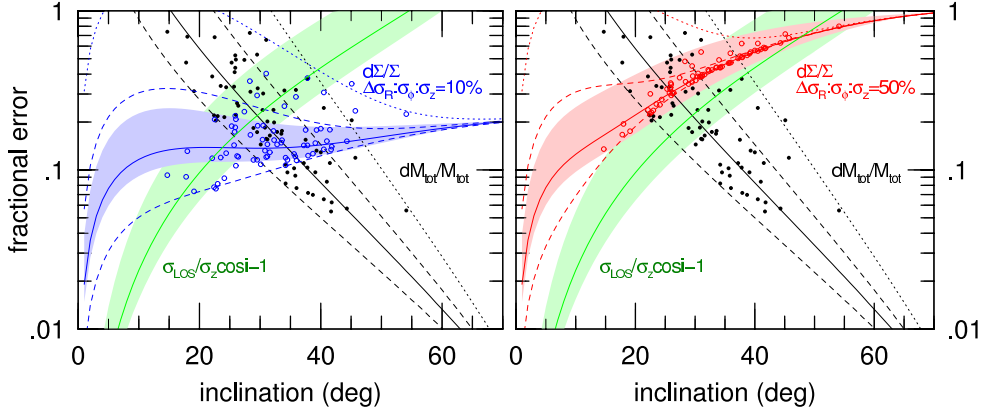}{1.9in}{-90}{50}{50}{-195}{220}
\caption{The Disk-Mass Survey error budget (fractional error) versus
inclination for total mass (dM$_{tot}$/M$_{tot}$) and disk mass
surface-density (d$\Sigma$/$\Sigma$) due to uncertainties in kinematic
inclinations and the stellar ellipsoid ratios.  Total mass errors
(black curves and dots) depend only on inclination uncertainties.
Disk mass surface-density errors depend also on the ellipsoid
decomposition; we demonstrate 10\% (left; gray [blue] curve
and points) and 50\% (right; gray [red] curve and points)
uncertainties in the ellipsoid decomposition. The latter are typical
of uncertainties reported in the literature. The
contribution of $\sigma_z$ to $\sigma_{\rm LOS}$ is shown for reference
(light gray [green]). This analysis shows our sample
inclination range of $30 \pm 15^\circ$ roughly balances disk and total
errors. Our ellipsoid work here shows we can achieve
ellipsoid-decomposition errors approximately as depicted in the
left-hand plot. Errors are for individual galaxies, which will be
reduced 3$\times$ for the survey of 40 galaxies sub-divided by color,
central surface-brightness, size, or luminosity.}
\end{figure}

Despite the proper treatment of the full 2D kinematic data in our current
analysis, isolation of the dynamical constraints is unsatisfactory,
i.e., providing inconsistent, and sometimes unphysical results with
large uncertainties. We find systematic effects in our fitting results
not yet explored in the literature. These may be due, in part, to the
assumed exponential form for $\sigma_R(R)$.  An assumed radial form for
$\sigma_R(R)$ is, indeed, required in isolating the AD equation for ellipsoid
decomposition due to the logarithmic derivative term (see Binney \& Tremaine
1987).  While strictly unnecessary in all other contexts of decomposing the
ellipsoid from the data, we have relied on a parametric approach to constrain
the allowed ellipsoid parameter space and reduce our random errors.
An exponential form for $\sigma_z(R)$ is
reasonable if light traces mass in galactic disks of constant scale
height; however, the same form for $\sigma_R(R)$ has no such
foundation despite its use in the literature.
Another concern is that the form of the dynamical equations is
inaccurate due to ignoring higher order terms in the moments of the
Boltzmann equation.  We have performed numerical orbital integrations
in a logarithmic potential to determine
$\sigma_{\phi}/\sigma_R$ directly for a range in input forms for
$\sigma_R(R)$ (Mihos, {\it private communication}).  In all cases, we
find the epicycle approximation to be a robust estimator for
$\sigma_{\phi}/\sigma_R$ beyond $\sim0.5h_R$.

\begin{figure}[!ht]
\plottwo{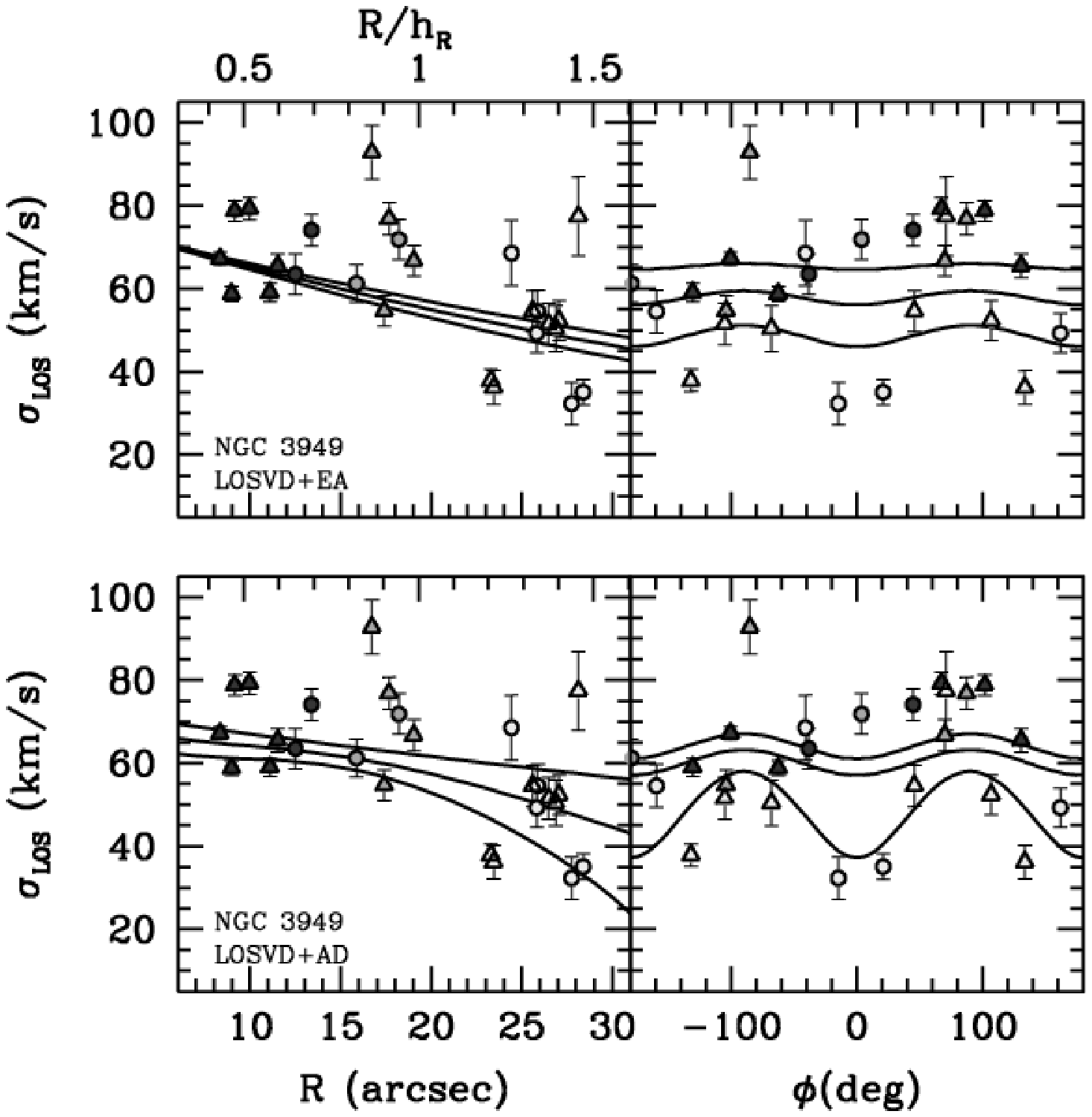}{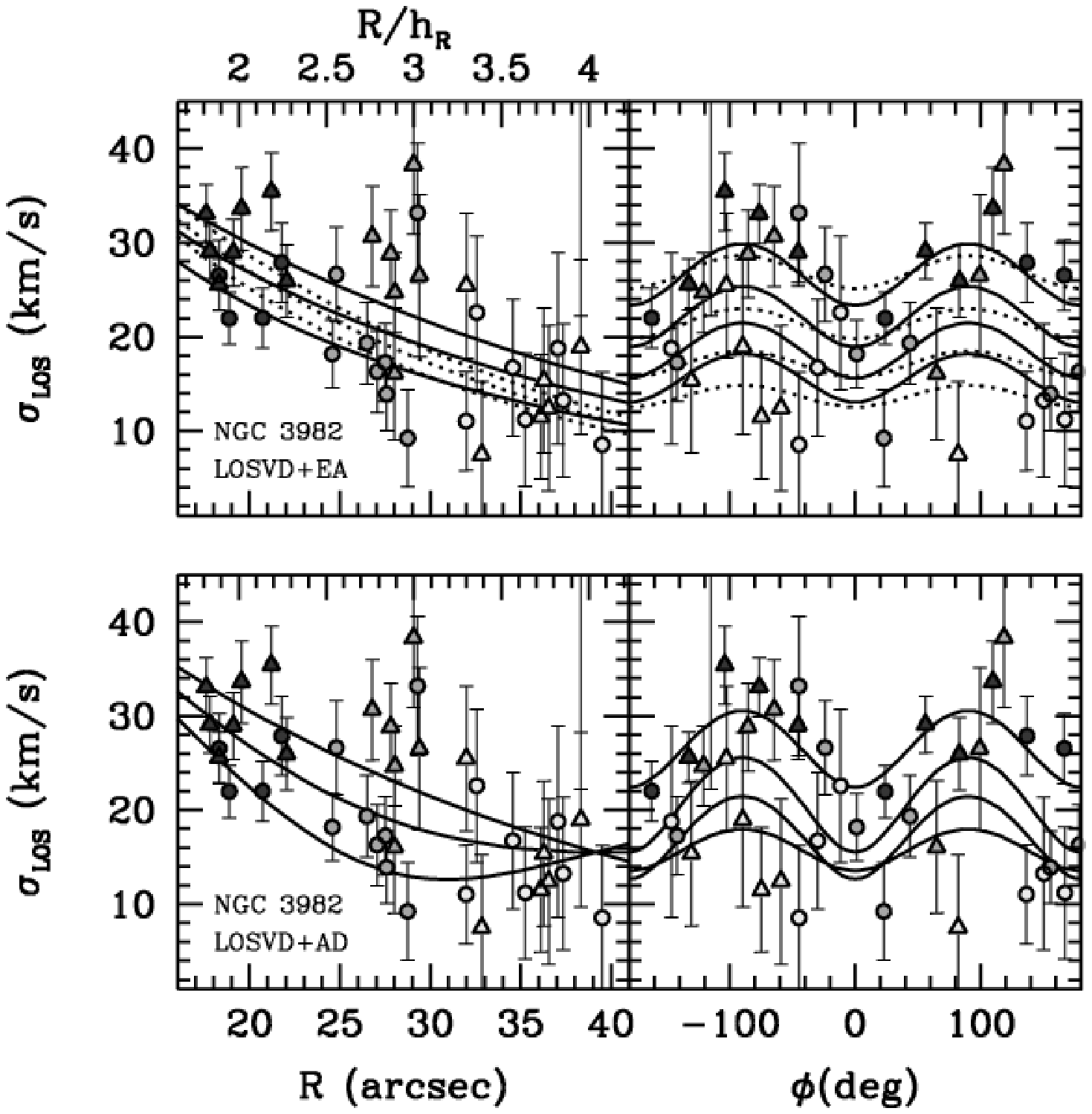}
\caption{Fits of the radial and azimuthal trends in $\sigma_{\rm LOS}$
for both NGC 3949 ({\it left}) and NGC 3982 ({\it right}) when using a
formalism incorporating only EA ({\it top}) or AD ({\it bottom}).
Best-fitting models are plotted in black ($\phi=0,45,90$ to the left
and $R=10,17,27$ for NGC 3949 and $R=20,25,30,35$ for NGC 3982 to the
right), and the point grayscale and type are coded by radius and
azimuth, respectively.  For the LOSVD+EA fit to NGC 3982, we plot an
additional fit to the data ({\it dotted line}) resulting from fixing
$\sigma_z$ to the result from the AD-only fit; the fitted $\sigma_R$
is within the errors of the AD-only fitted value.  For NGC 3949, we
find that the results from each dynamical assumption do not
necessarily reproduce the same results.  Non-physical results may be
due to incorrect assumptions including the variation of $\xi$ with $R$
from the AD equation produced by the assumption that $\sigma_R\propto
e^{-R/h_{\sigma}}$, where $h_{\sigma}$ is the e-folding length. }
\end{figure}

{\scriptsize
\begin{center}
\begin{tabular}{lrrrrrr}
\multicolumn{7}{c}{\small \bf Table 1: Fitted Velocity Field Parameters} \\
\hline\hline
 & \multicolumn{1}{c}{$i$} & \multicolumn{1}{c}{PA} & \multicolumn{1}{c}{$V_{{\rm rot},g}\ {\rm sin}i$} & \multicolumn{1}{c}{$h_{{\rm rot},g}$} & \multicolumn{1}{c}{$V_{{\rm rot},\star}\ {\rm sin}i$} & \multicolumn{1}{c}{$h_{{\rm rot},\star}$} \\
 Galaxy & \multicolumn{1}{c}{(deg)} & \multicolumn{1}{c}{(deg)} & \multicolumn{1}{c}{(km/s)} & \multicolumn{1}{c}{(arcsec)} & \multicolumn{1}{c}{(km/s)} & \multicolumn{1}{c}{(arcsec)} \\
\hline
NGC 3949 & 55.29$\pm$1.84 & 298.4$\pm$0.7 & 119.6$\pm$2.7 & 15.51$\pm$1.21 & 106.1$\pm$4.7 & 20.09$\pm$2.06 \\
NGC 3982 & 24.49$\pm$5.36 & 191.2$\pm$0.6 &  92.8$\pm$2.0 &  9.71$\pm$0.91 & 88.5$\pm$1.9 & 11.59$\pm$0.76 \\
\hline
\end{tabular}
\end{center}
}

{\footnotesize
\begin{center}
\begin{tabular}{llll}
\multicolumn{4}{c}{\small \bf Table 2: \boldmath $\sigma_z/\sigma_R$ Results} \\
\hline\hline
Method & Reference & \multicolumn{1}{c}{NGC 3949} & \multicolumn{1}{c}{NGC 3982} \\
\hline
Major Axis LOSVD+EA+AD & Westfall et al. (2007) & $1.18^{+0.36}_{-0.28}$ & $0.73^{+0.13}_{-0.11}$ \\
2D LOSVD + EA           & Current Work           & $1.12^{+3.51}_{-1.01}$ & $0.00^{+0.12}$         \\
2D LOSVD + AD           & Current Work           & $0.25^{+0.19}_{-0.25}$ & $0.46^{+0.03}_{-0.03}$ \\
2D LOSVD + EA + AD      & Westfall et al. (2008) & \nodata                & $0.62\pm0.20$ at 1h$_R$ \\
                        &                        & \nodata                & $0.31\pm0.05$ at 2h$_R$\\
\hline
\end{tabular}
\end{center}
}

With this assurance, our current methodology can be improved
dramatically by incorporating both EA and AD simultaneously and
solving the resulting ordinary differential equation in
$\sigma_R(R)$. This enables us to drop any assumptions about the
radial form of $\sigma_R(R)$, and, in turn, directly measure the gradient
in $\sigma_z/\sigma_R$.  Preliminary results from this new methodology
(Westfall et al. 2008, {\it in preparation}) are illustrated in Figure
3 for NGC 3982.  We find $\sigma_z/\sigma_R$ drops by a factor of two from 1 to 2
disk scale-lengths.  At 2 disk scale-lengths the error in the
ellipsoid is only 15\%, which bodes well for the DMS
error-budget.

\begin{figure}
\plotfiddle{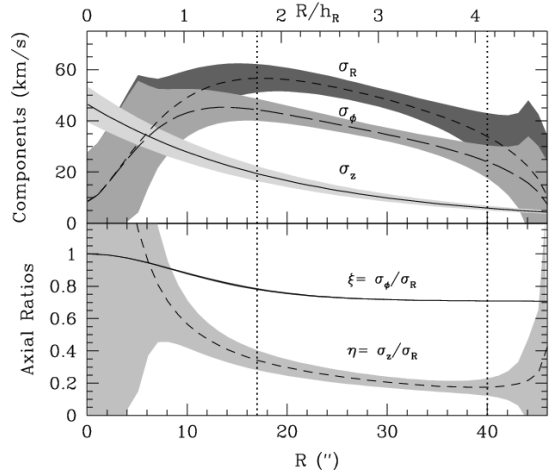}{2.1in}{0}{45}{45}{-110}{-10}
\caption{Decomposition of velocity ellipsoid of NGC 3982 using
dynamical constraints from EA and AD simultaneously and our 2D LOSVD
ionized gas ([{\scshape Oiii}]$\lambda 5007$\AA) and stellar
(Mg{\scshape i}b-triplet region; $\sim 4990-5250$\AA) kinematics.
The kinematics are fitted within the radial range delineated by the
dotted lines.  The ionized gas and stellar rotation curves are modeled with
independent hyperbolic tangent
functions, and the e-folding of $\sigma_z$ is taken to be 2$h_R$.  The shaded
regions show the formal random errors in the fitted model.
The resulting decomposition shows significant evidence for a
decrease in $\sigma_z/\sigma_R$, with a decomposition error of 15\% at
2 disk scale-lengths.}
\end{figure}

\acknowledgements We acknowledge financial support from the AAS
International Travel Grant and the National Science Foundation
(AST/0607516), and fruitful insights from P. C. van der Kruit,
K. Freeman, and C. Mihos.

\end{document}